\documentclass{article}

\def\mytitle#1{\setcounter{equation}{0}
\setcounter{footnote}{0}
\begin{flushleft}\Large\textbf{#1}\end{flushleft}
\vspace{0.25cm}}
\def\myname#1{\leftline{{\large #1}}\vspace{-0.13cm}}
\def\myplace#1#2{\small\begin{flushleft}\textit{#1}\\
\texttt{#2}\end{flushleft}}

\usepackage{graphicx}
\begin{document}

\mytitle{ACCRETION OF HOLOGRAPHIC DARK ENERGY : DEPENDENCY ONLY UPON HORIZON RADIUS OF EXPANDING UNIVERSE}

\frenchspacing \myname{ Ritabrata Biswas**,~Nairwita
Mazumder$\dag$,~Subenoy Chakraborty* } \frenchspacing

\myplace{Department of Mathematics, Jadavpur University,
Kolkata-32, India.}
{[**biswas.ritabrata@gmail.com~,~$\dag$nairwita15@gmail.com,*schakraborty@math.jdvu.ac.in]}


\begin{abstract}
In this paper we deal with accretion of dark energy in the holographic dark energy model for a general non-rotating static spherically symmetric black hole. The mass of the black hole increases or decreases depending on the nature of the holographic dark energy (quintessence or phantom) as well as on some integration parameters. It is to be illustrated that the enhancement or reduction of mass of a black hole is independent of the mass or size of the black hole itself. Rather it depends only upon the radius of the event horizon of the universe. Finally, the generalized second law of thermodynamics has been studied on the event horizon to be assured that the law holds even if when the black hole mass is decreasing though it is engrossing some mass.
\\\\
Key words: Black hole accretion, phantom energy, thermodynamics.
\end{abstract}

\section{Introduction}

It is popularly believed among the recent time astrophysicists and
the cosmologists that our universe is experiencing an accelerated
expansion. The strong supports of this idea of cosmic acceleration
come from recent observational data of type Ia supernovae
( Riess, A. G. et al. 1998; Perlmutter, S. et al. 1999;  Astier, P. et al. 2006) in associated with Large Scale
Structure (Tegmark, M. et al. 2004; Abazajian, K. et al. 2004, 2005) and Cosmic
Microwave Background anisotropies (Spergel, D. N. et al. 2003, 2006) have
provided main evidences. So
there was a plenty of requirements for theories to be constructed
which would be efficient enough to explain this interesting
behavior. It was seemed that this way of explaination has taken
two different paths namely introducing some modified gravity
theory, such as $f(R)$ gravity, where the extra terms in the
modified Friedmann equations are responsible for the acceleration.
The second way was to introduce some non-baryonic matter (known as
dark energy (DE)) having negative pressure(violating the strong
energy condition $\rho+3p>0$) in the framework of general
relativity. Though still the nature and cosmological origin of the
DE have remained enigmatic at present, one recent proposal is the
dynamical DE scenario (Copeland, E. J. 2006). The cosmological
constant puzzles may be better interpreted by assuming that the
vacuum energy is canceled to exactly zero by some unknown
mechanism and introducing a DE component with a dynamically
variable equation of state. From some scalar field mechanism which
suggests that the energy form with a negative pressure is provided
by a scalar field evolving under a suitable potential the
dynamical DE paradigm is often realized. Careful analysis of
cosmological observations, in particular of the WMAP(Wilkinson
Microwave Anisotropy probe) experiment (Spergel, D. N. et al. 2003; Bennett, C. et al. 2003;  Peiris, H. V. et al. 2003) indicates that the two-thirds of the total energy of our
universe is been occupied by the DE where as dark matter occupies
almost the rest. This DE is thought to be responsible for the
accelerated universe. Many models have been constructed for
interpreting this component. Cosmological constant or vacuum
energy (Weinberg, S.  1989; Carroll, S. M. 2004
; Peebles, P. J. E. et al  2003; Padmanabhan, T.  2003) and
quintessence models (Wetterich, C. 1988; Peebles, P. J. E.et.al.  1988; Ratra, B.et. al. 1988;  Frieman, J. A.et. al. 1995; Turner, M. S.et.al. 1997;  Caldwell, R. R.et. al. 1998; Liddle, A. R.et. al. 1999; Zlatev, I.et. al. 1999; Steinhardt, P. J.et. al. 1999; Torres, D. F.  2002)
are two very much popular models among these. However, as is well
known, there are two difficulties which arise from all of these
scenarios,$-$
the fine-tuning problem and the cosmic coincidence problem. The
fine-tuning problem asks why the DE density today is so small
compared to typical particle scales. The DE density of order
$10^{-47}GeV^{4}$, which appears to require the introduction of a
new mass scale. The second difficulty, the cosmic coincidence
problem, states : Since the energy densities of DE and dark mater
scale so differently during the expansion of the universe, why are
they nearly equal today? To get this coincidence, it appears that
their ratio must be set to a specific, infinitesimal value in the
very early universe. Recently, considerable interest has been
stimulated in explaining the observed DE by the Holographic
DE(HDE) model. For an effective field theory in a box of size $L$,
with UV cut-off $\Lambda_{c}$ the entropy $S$ scales extensively,
$S\sim L^{3}\Lambda_{c}^{3}$. However, the peculiar thermodynamics
of black
hole(BH) (Bekenstein, J. D. 1973,1974,1981,1994; Hawking, S. W.  1975,1976) has led Bekenstein to postulate that the maximum entropy
in a box of volume $L^{3}$ behaves nonextensively growing only as
the area of the box,i.e., there is a so called Bekenstein entropy
bound, $S\leq S_{BH} \equiv \pi M_{p}^{2}L^{2}$. This nonextinsive
scaling suggests that quantum field theory breaks down in large
volume. To reconcile this breakdown with the success of local
quantum field theory in describing observed particle
phenomenology, Cohen et. al. (1999) proposed a more
restrictive energy bound. They pointed out that in
quantum field theory a short distance (UV) cut off is related to a
long distance(IR)cut off due to the limit set by forming a black
hole. In the other words, if the quantum zero point energy density
$\rho$ is relevant to a UV cut-off, the total energy of
the whole system with size $L$ should not exceed the mass of a BH
of the same size, thus we have $L^{3}\rho_{\Lambda}\leq L
M_{p}^{2}$, this means that the maximum entropy is in order of
$S_{BH}^{\frac{3}{4}}$. When we take the whole universe into
account, the vacuum energy related to this holographic principle
(Hooft, G. T. 1993;  Susskind, L. 1994) is viewed as DE, usually dubbed HDE. The
largest IR cut-off L is chosen by saturating the inequality so
that we get the HDE density
\begin{equation}\label{1}
\rho=3c^{2}M_{p}^{2}L^{-2}
\end{equation}
where $M_{p}=\frac{1}{\sqrt{8\pi G}}$ is the reduced Planck mass.
c is a numerical constant. If we take L as the size of the current
universe for instance the Hubble scale $H^{-1}$, then the DE
density will be close to the observed data.

In HDE paradigm (Cohen et. al.  1999; Horava, P.et.al. 2000; Thomas, S. D. 2002;
 Hsu, S. D. H. 2004; Li, M. 2004;
 Pavon, D.et. al. 2005) one
determines an appropriate quantity to serve as an IR cut off
for the theory and imposes the constraint that the total vacuum
energy in the corresponding maximum value must not be greater than
the mass of a BH of the same size. By saturating the inequality
one identifies the acquired vacuum energy as HDE. Although the
choice of the IR cut off has raised on discussion in the
literature (Li, M. 2004; Pavon, D.et. al. 2005; Gong, Y. 2004; Guberina, B.et. al. 2005; Setare, M. R.  2007A, 2007B), it
has been shown, and it is generally accepted, that the radius of the event horizon of the universe($R_{h}$) the most suitable choice for the IR cut off where $R_{h}$ is defined as
(Hsu, S. D. H. 2004)
\begin{equation}\label{2}
R_{h}=a\int_{t}^{\infty}\frac{dt}{a}=a\int_{t}^{\infty}\frac{da}{Ha^{2}}
\end{equation}
which leads to results compatible with observations. Here $'a'$ is the scale factor of the background metric of the universe and H is the corresponding Hubble parameter.

 The
holographic energy density $\rho$ is then given by(taking $8\pi G=1$)
\begin{equation}\label{3}
\rho=\frac{3c^{2}}{R_{h}^{2}},
\end{equation}
Furthermore, we can define the dimensionless dark energy as:
\begin{equation}\label{4}
\Omega\equiv \frac{\rho}{3H^{2}}=\frac{c^{2}}{R_{h}^{2}H^{2}}
\end{equation}
In the case of a dark-energy dominated universe, dark energy
evolves according to the conservation law
\begin{equation}\label{5}
\dot{\rho}+3H(p+\rho)=0
\end{equation}
or equivalently:
\begin{equation}\label{6}
\dot{\Omega}=H\Omega\left(1-\Omega\right)\left(1+\frac{2\sqrt{\Omega}}{c}\right)
\end{equation}
where the equation of state is
\begin{equation}\label{7}
p=\omega_{D}\rho
\end{equation}
and consequently the index of the equation of state is
of the form:
\begin{equation}\label{8}
\omega_{D}=-\frac{1}{3}\left(1+\frac{2\sqrt{\Omega}}{c}\right)
\end{equation}
As we can clearly see, $\omega$ depends on the parameter c. In
recent observational studies, different groups have ascribed different
values to c. A direct fit of the present available SNe Ia data
indicates that the best fit result is the best-fit value $c =
0.21$ within $1-\sigma$ error range (Huang, Q. G.et. al.  2004A). In addition,
observational data from the X-ray gas mass fraction of galaxy
clusters lead to $c = 0.61$ within $1-\sigma$ ( Chang, Z. et. al. 2006).
Similarly, combining data from type Ia supernovae, cosmic
microwave background radiation and large scale structure give the
best-fit value $c = 0.91$ within $1-\sigma$ (Kao, H. C. et. al. 2005; Zhang, X. et. al. 2005, 2007), while combining data from type Ia supernovae, X-ray gas
and baryon acoustic oscillation lead to c = 0.73 as a best-fit
value within $1-\sigma$ (Wu, Q. et. al.  2007; Ma,Y-Z. et. al. 2007). Finally, the study of the
constraints on the dark energy arising from the holographic
connection to the small l CMB suppression, reveals that $c = 2.1$
within $1-\sigma$ error (Shen,J. et. al. 2004). In conclusion, $0.21 \leq c
\leq 2.1$, and HDE provides the mechanism for the $w =0 .1$
crossing and the transition to the accelerating expansion of the
Universe.

In nature, the compact objects, particularly the black holes
(BHs), are not visible but can be detected by the presence of the
accretion disc around them. By analyzing  light rays off an
accretion disc, one can speculate the properties of the central
compact object.

Although the accretion phenomena around compact objects
(particularly BHs) have been extensively discussed over the last
three decades (e.g. Mukhopadhyay, B. 2003), it was started long ago
in 1952 by Bondi  (1952). He studied stationary spherical
accretion problem by introducing formal fluid dynamical equations
in the Newtonian framework. In the framework of general
relativity, the study of accretion was initiated by Michel
(1972). By choosing the Newtonian gravitational potential,
Shakura and Sunyaev (1973) formulated very simplistic but
effective model of the accretion disc. Some aspects of the
accretion disc in fully relativistic framework had been studied by
Novikov and Thorne (1973) and Page and Thorne (1974).
Subsequently, various aspects related to the critical behavior of
general relativistic flows in spherical symmetry have been studied
(Begelman, M. C. 1978; Brinkmann, W.  1980; Malec, E.  1999; Das, T. K. 2001). Although there are a few
steps forward, still it is extremely difficult to simulate the
full scale realistic accretion discs including outflows in a full
general relativistic framework.

 One may note that in BH accretion, an important issue is that the flow
 of accreting matter must be transonic in nature, i.e., there should be sonic
 point(s) (Chakrabarti, S. K. 1990, 1996A, 1996B) in the flow.
 On the other hand, accretion flow around a neutron star is not necessarily transonic (i.e., sonic point may or may not exist).

Due to present accelerated expansion of the universe, the matter
in the universe is dominated by DE (almost $\frac{2}{3}$ of the
matter is in the form of DE). Therefore it is reasonable to assume
the accreting matter is in the form of DE. Babichev et.al.
(Babichev, E. et. al. 2004,2006) were the pioneers to think about the
DE accretion upon a BH, in the framework of Bondi accretion
(Bondi, H. 1952).

In this paper we will study the accretion of HDE upon BH where in section 2 we will calculate the expression for the mass change of the BH. Section 3 contains the thermodynamical analysis. Finally in the chapter 4 we will discuss the whole thing derived in the previous sections.
\section{Equation governing the accretion of HDE on general static non rotating BH}
In this section, we shall calculate the rate of mass accretion
using the conservation equations in fluid dynamics. For that let
us consider the spherical accretion of HDE onto BH. For
simplicity, we assume a non-rotating spherically symmetric BH having metric ansatz
\begin{equation}\label{9}
ds^{2}=-h(r)dt^{2}+\frac{dr^{2}}{h(r)}+r^{2}d\Omega^{2}
\end{equation}
The lapse function $g_{00}=h(r)$ identifies the event horizon
$(r_{e})$ by setting $h(r)=0$ (i.e., $h(r_{e})=0$).

Suppose  the
HDE is represented by a perfect fluid for which the energy
momentum tensor be:
\begin{equation}\label{10}
T_{\mu\nu}=\left(\rho+p\right)u_{\mu}u_{\nu}+p g_{\mu\nu}
\end{equation}
where, $\rho$ and $p$ are respectively the energy density and
pressure of the HDE and $u^{\mu}=\frac{dx^{\mu}}{ds}$ is the four
velocity of the flow. Assuming the fluid flow in the radial
direction to be $u$ (note that $u<0$ as the fluid flow towards the
BH).  The explicit form of the components of $u^{\mu}$ is :
$$(u^{0},~u^{1},~0,~0)$$
It is to be noted that the third and the fourth component of
$u^{\mu}$ are zero due to spherical symmetry. Using the
normalization rule $u^{\mu}u_{\mu}=-1$, we have,
$$u_{0}=-\sqrt{h\left(r\right)+u^{2}}~~~~~~,~~~~~~ u_{1}=\frac{u}{h\left(r\right)}$$
\begin{equation}\label{11}
u^{0}=\frac{\sqrt{h\left(r\right)+u^{2}}}{h\left(r\right)}~~~~~~,~~~~~~
u^{1}=u
\end{equation}
As a consequence the explicit components of $T_{\mu\nu}$ (from
(\ref{10}))are given by,
$$T^{0}_{0}=-\rho-\frac{u^{2}}{h\left(r\right)}\left(\rho+p\right)$$
$$T^{1}_{1}=p+\frac{u^{2}}{h\left(r\right)}\left(\rho+p\right)$$
So, we have got the stress energy tensor components and the velocity components. Now to determine different dynamical parameters we need to construct and solve different differential equations. These will be provided from the conservation relations.
\begin{equation}\label{12}
T^{2}_{2}=p
\end{equation}
$$T^{3}_{3}=p$$
$$T^{1}_{0}=u\left(\rho+p\right)\sqrt{h\left(r\right)+u^{2}}$$
In the present problem we have two conservation relations namely,

{\bf(I) Conservation of mass flux : }

\begin{equation}\label{13}
J^{a}{; a}=0,
\end{equation}
 where $J^{a}$ is the current density. Explicitly it gives (after integrating with respect to 'r')
\begin{equation}\label{14}
\rho u r^{2}=\lambda
\end{equation}
with $\lambda$, an arbitrary constant  of integration. We have to
note that as $u<0$ so $\lambda$ is also $<0$.

{\bf(II) Energy-momentum conservation relation : }

\begin{equation}\label{15}
T_{\mu ; \nu}^{\nu}=0.
\end{equation}
In particular $T_{0; \nu}^{\nu}=0$ characterizes the energy flux
across the horizon. A first integral (w.r.t. 'r') of
$T_{0;\nu}^{\nu}=0$ gives
\begin{equation}\label{16}
\left(\rho+p\right)\left(h\left(r\right)+u^{2}\right)^{\frac{1}{2}}ur^{2}=a_{1}
\end{equation}
here $a_{1}$ is the arbitrary constant of integration.

We have three variables, namely $u$, $\rho$ and $p$ (the EoS parameter $\omega_{D}$ rather). Now as we got two integration of motion only still we are underdetermined.

We can have another integral of motion by projecting the
energy-momentum conservation relation (7) along the four velocity
$u_{\mu}$, i.e., $u^{\mu}T^{\nu}_{\mu;\nu}=0$ or in explicit form
$$u^{\mu}\rho_{,\mu}+\left(\rho+p\right)u^{\mu}_{;\mu}=0$$
Integrating once gives (Babichev, E. et. al. 2004)
\begin{equation}\label{17}
u
r^{2}exp\left\{\int^{\rho}_{\rho_{\infty}}\frac{d\rho^{'}}{\rho^{'}+p\left(\rho^{'}\right)}\right\}=-b_{1}
\end{equation}
This is also known as energy-flux equation. Here for convenience
the negative sign is chosen in front of the constant $'b_{1}'$.

Now removing $ur^{2} from  $(\ref{16}) and (\ref{17}) we obtain
\begin{equation}\label{18}
\left(\rho+p\right)\left(h\left(r\right)+u^{2}\right)^{\frac{1}{2}}exp\left\{-\int^{\rho}_{\rho_{\infty}}\frac{d\rho^{'}}{\rho^{'}+p\left(\rho^{'}\right)}\right\}=c_{1}
\end{equation}
with
$c_{1}=-\frac{a_{1}}{b_{1}}=\rho_{\infty}+p\left(\rho_{\infty}\right)$.
Now from the relations (\ref{17}) and (\ref{18}), the fluid
velocity $u_{H}$ and the density $\rho_{H}$ at the event horizon
are related by the relation (Babichev, E. et. al. 2004) (with $r_{e}=1$)
\begin{equation}\label{19}
b_{1}\frac{\rho_{H}+p\left(\rho_{H}\right)}{\rho_{\infty}+p\left(\rho_{\infty}\right)}=-\frac{b_{1}^{2}}{u_{H}^{2}}=exp\left\{2\int^{\rho_{H}}_{\rho_{\infty}}\frac{d\rho^{'}}{\rho^{'}+p\left(\rho^{'}\right)}\right\}
\end{equation}
Now eliminating $ur^{2}$from equations (\ref{14}) and (\ref{16})
we have
\begin{equation}\label{20}
\frac{\left(\rho+p\left(\rho\right)\right)}{\rho}\sqrt{u^{2}+h\left(r\right)}=\frac{a_{1}}{\lambda}
\end{equation}
Taking $r$ as the independent variable if we take differentials of
eq. (\ref{14}) and (\ref{20}) and eliminate $d\rho$ from them,
then after simplification we obtain
\begin{equation}\label{21}
\frac{du}{u}\left[-c_{s}^{2}+\frac{u^{2}}{h\left(r\right)+u^{2}}\right]+\frac{dr}{r}\left[-2c_{s}^{2}+\frac{1}{2}\frac{rh'(r)}{h\left(r\right)+u^{2}}\right]=0
\end{equation}
where, $1+c_{s}^{2}=\frac{d\ln\left(\rho+p\right)}{d\ln\rho}$.
Here the solution will be feasible if it passes through a critical
point and it characterizes the fluid falling into the BH with
monotonically increasing velocity. However the critical point
corresponds to vanishing of both the square brackets in equation
(\ref{21}) and the parameters at the critical point ($r_{*}$) as
(Babichev, E. et. al. 2004).
\begin{equation}\label{22}
u_{*}^{2}=\frac{1}{4}r_{*}h'(r_{*})~~~~~~,~~~~~~c_{s*}^{2}=\frac{u_{*}^{2}}{h(r_{*})+u_{*}^{2}}
\end{equation}
Further due to fluid accretion the rate of change of BH mass
(Huang, Q. G. 2004B) $\dot{M}=-4\pi r^{2} T_{0}^{1}$ can be written as
\begin{equation}\label{23}
\dot{M}=4\pi
b_{1}\left(\rho_{\infty}+p\left(\rho_{\infty}\right)\right)
\end{equation}
It is to be noted that the above equation is independent of the
mass of the BH (contrary to the Schwarzschild and
Reissner-Nordstr$\ddot{o}$m BHs
(Jamil, M. et. al. 2008; Gonzalez-Diaz, P.F. 2004; Cai R-G.et. al.  2006; Jimnez Madrid,J. A.et. al. 2008; Zhang, X.  2009; Guariento, D.C.et al 2008)). When the mass of the BH
concerned is not present or the expression is true for any
arbitrary BH mass we will not confine our system at  a infinite
distance only(as it is not dependent upon the potential itself
which is a function of BH mass). So equation (\ref{23}) holds at any finite point. So the above equation can be written for
any general $\rho$ and $p$ as done in  (Jamil, M. et. al. 2008, 2011A, 2011B)(satisfying the holographic equation of state and violating
weak energy condition), i.e.,
\begin{equation}\label{24}
\dot{M}=4\pi b_{1}\left(\rho+p\left(\rho\right)\right)
\end{equation}
When BH accretes fluid simultaneously it also radiates energy known as Hawking Radiation. This radiation (Saskind, L. 1992) causes the evaporation of the BH which is balanced by the accretion of matter into the BH and as a result the total system is supposed to be under equilibrium. But when we analyze the parameters, for example temperature of the accreting fluid at very far from the BH with very near to the BH there will be a huge difference. But in local cells the parameters show equilibrium nature. Such an equilibrium is called quasi equilibrium. In this paper we have considered large BHs in general. Now if we think about small BHs then as the Hawking temperature of the BH, $T=\left(8\pi M\right)^{-1}$ is inversely proportional to its mass, temperature will increase for the small BHs. This will cause more radiation according to the standard fourth order rule of black body radiation. under such huge amount of Hawking radiation the accretion-radiation equilibrium may not be the equilibrium one. In such a case we will not be able to speculate whether the accretion procedure is at all independent of the mass or not. For very small BHs the procedure of accretion is still a fact be research with. 

From equation (\ref{20}) we have (with $r_{e}=1$) the index of equation of state as
$$\omega_{D}=-1+\frac{a_{1}}{\lambda \left\{u^{2}+h\left(r \right)\right\}^{\frac{1}{2}}}$$
Note that $\omega_{D}~~>~~or~~<-1$ depends on the sign of the constant
$a_{1}$, i.e., the HDE is of phantom nature or not will depend
crucially on $a_{1}~~<~~or~~>0$.

\section{Thermodynamical analysis of accreting matter}
We shall now discuss the thermodynamics of the DE accretion that
crosses the event horizon of the BH given by equation
$p=\omega_{D}\rho$. Actually, we have two way motives for doing
this thermodynamical analysis. The first, is to determine the
value of $b_{1}$ such that we can determine the sign of $\dot{M}$
and second one is to examine the validity of the generalized
second law of thermodynamics which is an invariant law and to find
any restriction on the equation of state $\omega_D$ from
thermodynamical point of view. Let us rewrite the BH metric in the
form
\begin{equation}\label{25}
ds^{2}=h_{ab}dx^{a}dx^{b}+r^{2}d\Omega^{2}
\end{equation}
where $(a,~b)=(0,~1)$ and $h_{ab}=diag\left(-h(r),
\frac{1}{h(r)}\right)$.

For thermodynamical analysis we start with the work density $(W)$
and energy supply vector $\psi_{a}$ which are defined as
(Cai, R. G. et. al. 2007, Chakraborty, S.et. al.  2010)
\begin{equation}\label{26}
W=-\frac{1}{2}Trace~T_{a}^{b}=\frac{1}{2}\left(\rho-p\right)
\end{equation}
and
\begin{equation}\label{27}
\psi_{a}=T_{a}^{b}\partial_{b}r+W\partial_{a}r
\end{equation}
i.e.,
$$\psi_{0}=T_{0}^{1}=-u\left(\rho+p\right)\sqrt{u^{2}+h\left(r\right)}$$
and
$$\psi_{1}=T_{1}^{1}+W=\left(\rho+p\right)\left(\frac{1}{2}+\frac{u^{2}}{h(r)}\right)$$
Where $T_{ab}$ is the projected energy-momentum tensor, normal to
the 2-sphere. Then the change of energy across the event horizon
is given by (Chakraborty, S.et. al.  2010)
$$-dE\equiv -A\Psi=-A\left[\psi_{0}dt+\psi_{1}dr\right]$$
Hence the energy crossing the event horizon is (Chakraborty, S.et. al.  2010;
Mazumder, N. et. al. 2009) (choosing $r_e=1$)
\begin{equation}\label{28}
dE=4\pi u^{2}\left(\rho+p\right)dt
\end{equation}
so comparing equations (\ref{24}) and (\ref{28}) (as $E=mc^{2}$
and $c=1$) the arbitrary constant $b_{1}$ is given by
\begin{equation}\label{29}
b_{1}=u^{2}
\end{equation}
i.e.,
\begin{equation}\label{30}
\dot{M}=4\pi u^{2}\left(\rho+p\right)
\end{equation}
So we can say that $\dot{M}>0$ in quintessence era,i.e., the BH
mass is increasing there though the rate of increasing is slowing
down as we move towards the phantom barrier line. While in phantom
era $\dot{M}<0$,i.e., the mass of the BH is decreasing.
Where the value of $\dot{M}$ starts to decrease is a point of
interest. To calculate that we will use (\ref{3}) and (\ref{8}) in
(\ref{30}) and with the help of (\ref{4}) we have
$$\dot{M}=8\pi
u^{2}\frac{c^{2}}{R_{h}^{2}}\left(1-\frac{1}{R_{h}H}\right)$$
which on differentiation gives 
\begin{equation}\label{30}
\frac{d\dot{M}}{dR_{h}}=8\pi
u^{2}\frac{c^{2}}{R_{h}^{4}}\left(\frac{3}{H}-2R_{h}\right)
\end{equation}
So, $\dot{M}$ increases when $R_{h}<\frac{3}{2}R_{H}$ where
$R_{H}$ is the Hubble radius , $R_{h}$ is the radius of the event horizon (defined in (\ref{2})) and $\dot{M}$ decreases when
$R_{h}>\frac{3}{2}R_{H}$.

\begin{figure}
\includegraphics[height=2.5in, width=5in]{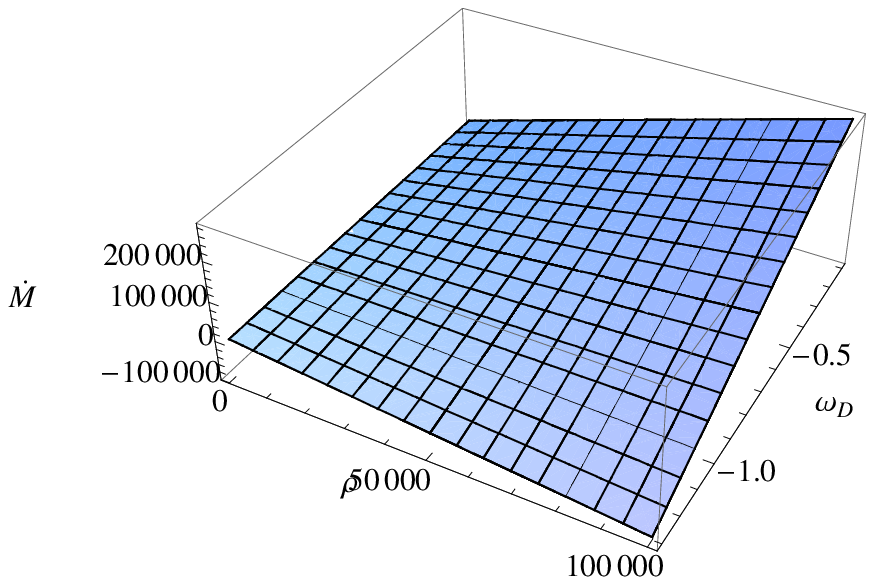}$~~~~~~~~~~~~~~~~~~~~~~~~~~~~~~~~~~~$\\
$~~~~~~~~~~~~~~~~~~~~~~~~~~~~~~~~~~~~~~~~~~~~~~~~~~~~~~~~~~~~$Fig.1.

Here we have plotted $\dot{M}$ with the variation of $\omega_{D}$
and $\rho$. \hspace{1cm} \vspace{2cm}
\end{figure}
In fig.1. we have plotted $\dot{M}$ with the variation of
$\omega_{D}$ and $\rho$. This is not a graph to be scaled
actually. The graph shows that
$\dot{M}$ is negative when $\omega_{D}<-1$.

 Now it is quite awkward to hear that the BH is absorbing
fluid but the mass of the BH is decreasing. Obviously the negative pressure of the dark energy
model is responsible for this incidence. We shall now examine the validity of the generalised second law of thermodynamics in the present case. Using Clausius relation, the time
variation of horizon entropy is given by
\begin{equation}\label{31}
\dot{S}_{h}=\frac{4\pi u^{2}}{T_{e}}\left(\rho+p\right)
\end{equation}
where $T_{e}~~(=\left|\frac{f'(r)}{4\pi}\right|_{r=r_{e}})$ is the
temperature of the event horizon. We now study the entropy
variation of the matter in the form of HDE bounded by the event
horizon. We assume that the thermodynamical system bounded by the
event horizon is an equilibrium one and hence the temperature of
the matter inside the event horizon is same as $T_{e}$. From the
Gibb's equation
\begin{equation}\label{32}
T_{e}dS_{I}=dE+pdV
\end{equation}
we obtain
\begin{equation}\label{33}
T_{e}\dot{S}_{I}=4\pi
u\left(\rho+p\right)\left[\sqrt{h(r)+u^{2}}+\frac{up}{\rho}\right]
\end{equation}
Thus the total entropy change is given by
\begin{equation}\label{34}
\left(\dot{S}_{h}+\dot{S}_{I}\right)=\frac{16\pi^{2}u^{2}
\rho\left(1+\omega_{D}\right)}{|h'(r)|}\left[1+\omega_{D}+\sqrt{1+\frac{h(r)}{u^{2}}}\right]
\end{equation}
From the above expression of the total entropy change (i.e., eq.
(\ref{34})) we see that the term outside the square bracket in the
right hand side is positive in phantom era. So to make the time variation of the total entropy
to be positive the square bracket must be a positive quantity also
and it is true again in the phantom era. So GSLT is valid in
the phantom era. But in quintessence era the term out side the
square bracket is negative. So the square bracket must be negative
to ensure the validity of the GSLT, which lead us to a lower
boundary for the equation of state parameter as
$$\omega_{D}>-1+\sqrt{1+\frac{h(r)}{u^{2}}}.$$

\section{Conclusion  :}

The work deals with the possibility of DE accretion onto a general
static spherically symmetric BH. The analysis shows that the
evolution of mass of a static BH does not depend on its mass
rather it depends on the equation of state of the HDE. Due to
static non rotating BH the accreting DE falls radially ($u<0$) on
the BH. Although, the accreting HDE satisfies GSLT but the BH
evaporates in phantom era while in quintessence era BH only
accretes but the equation of state must have an lower bound for
the validity of GSLT. The figure shows the variation of $\dot{M}$ against energy density $\rho$ and equation of state parameter $\omega_{D}$. As expected we see that in phantom era whatever be the value of $\rho$, $\dot{M}$ is negative while in quintessence era $\dot{M}$ is positive. Further, the variation of the BH mass does not depend on any BH parameters, rather it depends on the horizons of the space time. Therefore, we may conclude that the geometry of the background space-time has a signified effect on the BH accretion of DE.For further work it will be interesting to
study the accretion of HDE for rotating BH and examine the
influence of
rotation on the BH evaporation.\\

{\bf Acknowledgement :}

 RB
wants to thank West Bengal State Government for awarding JRF. NM
wants to thank CSIR, India for awarding JRF. SC is thankful to DST-PURSE Programme, Jadavpur University.
All the authors are
thankful to IUCAA, Pune as this work was done during a visit.

\end{document}